\documentclass[aps,pre,preprint,floatfix,twoside,tightenlines,showpacs,
showkeys]{revtex4}
\usepackage{amsmath,graphicx}
\begin{document}
%
\title{Generalized Lattice Model of Multi-Component Systems with Internal Degrees of Freedom. I. General consideration}
\author{A.Yu.~Zakharov}\email[E-mail: ]{Anatoly.Zakharov@novsu.ru}
\author{M.I.~Bichurin}\email[E-mail: ]{Mirza.Bichurin@novsu.ru}
\affiliation{Novgorod State University, Veliky Novgorod, 173003, Russia}
%
%
\begin{abstract}
The paper contains the generalization of usual lattice model of multicomponent systems. The generalization is related to account the following factors: 1. The short-range parts of interatomic repulsions. These repulsions are not identical for different pairs of atoms, therefore it is impossible to take into account the repulsions by means of usual ideal lattice introduction. 2. The long-range interatomic potentials take into account by means of effective fields approximation. 3. The presence the interatomic potentials depending on some inner degrees of freedoms such as atomic electric and/or magnetic momentum. The Helmholtz free energy functional in the generalized lattice model is reduced to the Ginzburg-Landau-Cahn-Hilliard-like (GLCH) form. The connection between the interatomic potentials characteristics and the parameters of the GLCH-like functional is obtained. The equations for both full and partial equilibrium distributions of the species in multicomponent systems are derived.
\end{abstract}
%
\maketitle

\section{Introduction}

The problem of quantitative description of multicomponent systems is interesting from both theoretical and practical points of view. The theoretical description of real systems from ``first principles'' is now impossible because of absence of constructive methods of partition functions calculations for realistic models. Therefore the only way of the realistic models investigations is the phenomenological approach based on some additional assumptions. One of the most popular phenomenological approaches to statistical thermodynamics of condensed systems is the Cahn-Hilliard and Ginzburg-Landau approximation. Application of these very attractive models has some well known restrictions. Most essential of these restrictions  is absence of concrete relations between models parameters and characteristics of constituents and their interactions.

At construction of the theory of these structures there is a lot of the difficulties caused by absence of constructive methods of the analysis of realistic models of substance from ``first principles''. Therefore it is necessary to be content with development of phenomenological models. The examples of such models are Ginzburg-Landau~\cite{LL}, Cahn-Hilliard~\cite{Cahn1, Cahn2, Cahn3} models and the various variants of lattice and cell models~\cite{Khach, Smir,Israel,McCoy}.

The basic idea of lattice models consists on introduction of some ideal lattice on which sites atoms of components are distributed. This idea allows to consider the identical short-range repulsions between atoms, but does not allow to consider differences in repulsions of the components.
Most essential disadvantages of the lattice models are well known:
\begin{enumerate}
    \item The assumption of some lattice existing independently of
    difference between components proper atomic sizes. This assumption is
    comparatively plausible by the condition of vanishingly small
    difference of the components atomic sizes. As a rule, this condition
    takes no place. Really, introduction of lattice is the way the
    short-range parts of interatomic potentials account. Difference of
    short-range parts of components interatomic potentials leads to
    lattice distortions and the notion of the lattice, strictly
    speaking, in this case becomes invalid.
    \item The postulate on Fermi-like functional form of distribution
    function for average occupations numbers has not any theoretical
    justification~--- there are infinitely many functions with range
    of values $[0;1]$.
    \item Interatomic potentials in lattice-like models of
    condensed systems, as a rule, cannot  be described with
    small number of parameters, such as interaction energy
    of nearest neighbors interactions. The unavoidable
    lattice  distortions lead to the changes of interatomic distances
    and short-range parts of interatomic potentials are not slowly
    changing functions.
    \item The assumption on pair-wise interatomic potentials is not
    very convincing, because all the interatomic interactions appear
    as result of the averaging over fast electronic degrees of
    freedom in system. These degrees of freedom are very sensitive to
    atom coordination environment.
\end{enumerate}

Ginzburg-Landau and Cahn-Hilliard models  contain a set of phenomenological parameters which connections with real physical characteristics of the components and their interactions remains hidden.

The generalized lattice model (GLM) of multicomponent condensed systems (such as solid or liquid solutions) was proposed in papers~\cite{ZT1,ZT2,ZT3} and developed in~\cite{ZL,Z-2008,ZZUO}. In contradistinction to usual
lattice model (see for example~\cite{Khach,Smir,Israel,McCoy}), the GLM takes into account the following factors:
\begin{enumerate}
    \item The short-range interatomic repulsions~\cite{ZT1}. These repulsions
    are not identical for different pair of atoms, therefore it is
    impossible to take into account the repulsions by means of
    lattice introduction (at best, the lattice model can be considered
    as method of short-range interactions account in the case of
    one-component system and that with some essential restriction,
    related  in particular to thermal defects).
    \item The presence of the local fields due to the long-range parts of the
    interatomic potentials~\cite{ZT2}. These fields have the essential
influence on both equilibrium properties~\cite{ZT1} and non-equilibrium processes~\cite{ZT3} on the corresponding scales.
    \item The connection between the GLM and the Ginzburg-Landau approximation is obtained in paper~\cite{ZL}. The relation of some characteristics of the components and their interatomic potentials with Ginzburg-Landau parameters is established. This relation makes possible to use the mathematical tools of the Ginzburg-Landau and Cahn-Hilliard approximations for the GLM research.

    \item The existence of comparatively stable polyatomic complexes
    that manifest itself in both thermodynamics and kinetics as one
    and indivisible particles~\cite{ZZUO}.
\end{enumerate}

The present paper contains further development of the GLM. In addition
to the previous results it takes into account the following factors:
\begin{enumerate}
	\item Existence of the internal atomic degrees of freedom such as atomic electric and magnetic moments. These degrees of freedom are responsible for the local magnetization and local electric polarization in the system.
    \item Reduction of the GLM to the Ginzburg-Landau-Cahn-Hilliard-like (GLCH) functional form. In contrast to the GLCH theory, this functional contains the well defined parameters that have direct connections with the  characteristics of the constituents and their interactions.
    \item Colossal times of the structure transformation and the
    equilibrium reaching. Real condensed systems are, as a rule,
    essentially non-equilibrium systems.
    \item The presence of the hierarchy of relaxation times in real
    condensed systems and related partially equilibrium states of the
    systems. The evolutional processes rearrangements in such systems
    have multi-stepped character.
    \item Absence of full thermodynamic equilibrium in the system.
\end{enumerate}

All the results are formulated on the basis of the unified
mathematical apparatus and the common physical ideas. All the
approximations have clear physical sense, well foundations, and
strictly based conditions of the applicability.

\section{Generalized lattice model --- the basic notions and relations}

\subsection{Conditions for free energy}
The basic idea of the lattice models is the assumption that particles are located  in sites of some ideal lattice. This assumption is incompatible with differences of atomic sizes of the components and with presence of the various kinds defects in real condensed systems.

Therefore let us introduce instead of the ideal lattice the connections between local densities of the components with a view to take into account the differences of the atomic sizes of the components and the various kinds of defects in the system.

It should be noted that the short-range repulsive parts of the interatomic potentials lead, in particular, to some restriction on the local densities  $n_i(\mathbf{r})$ of the constituents particles numbers in the system
\begin{equation}\label{dens}
    n_i(\mathbf{r}) \leq \frac{1}{\omega_i},
\end{equation}
where $\omega_i$ is the inverse value of the maximal local density of $i$-th component ($i=1\div m$, $m$~is the number of the components in the system). The quantity $\omega_i$ has dimensionality of volume and henceforth will called as the specific atomic volume of $i$-th component.
As far as the quantity $\omega_i\, n_i(\mathbf{r})$ is the local volume fraction of $i$-th component at the point $\mathbf{r}$, then we have the restrictions on the local densities of the components for all points in the system:
\begin{equation}\label{pack-0}
    \sum_{i=1}^m\ \omega_i\ n_i(\mathbf{r}) \leq 1.
\end{equation}

Let us introduce the additional component --- the vacancies (or holes) with their proper volume $\omega_0$ and local density $n_0(\mathbf{r})$.
Suppose the holes do not interact with real components but they fill all the unavoidably existent vacant places in the system (such as thermal or radiation defects). With account of the vacancies we have the following {\em the packing condition}:
\begin{equation}\label{packing}
    \sum_{i=0}^m\ \omega_i\ n_i(\mathbf{r})-1=0.
\end{equation}
%
%
The minimization of free energy should be realized under the packing condition, which takes into account the short-range parts of the interatomic potentials in the system. Hence the interatomic potentials should be included into the free energy with cutting out their the short-range parts:
\begin{equation}\label{long-range}
    K_{ij}(\mathbf{r})=\left\{%
\begin{array}{ll}
    W_{ij}(\mathbf{r}), & \hbox{if} \  |\mathbf{r}|\geq a_{ij}, \\
    0, & \hbox{otherwise,} \\
\end{array}%
\right.
\end{equation}
where  $W_{ij}(\mathbf{r})$ is ``true'' interaction potential between $i$-th and  $j$-th components, $a_{ij}$ are the cutting parameters, related to the specific atomic volumes of the components by the relations
\begin{equation}\label{cut}
    a_{ij}\simeq \left[ (\omega_i)^{1/3}+
(\omega_j)^{1/3}\right].
\end{equation}

Beyond the packing condition~(\ref{packing}) the numbers $N_i$ of components atoms in the system should be fixed at the free energy minimization. These conditions have the following form:
\begin{equation}\label{N}
    \int\limits_{(V)}\ n_i(\mathbf{r})\ d\mathbf{r}-N_i=0, \qquad i=0\div m.
\end{equation}
The packing condition~(\ref{packing}) and condition of numbers particles conservation~(\ref{N}) should be satisfied for any form of the configuration part of the free energy.

\subsection{Configuration part of free energy}
Denote by $K_{ij}\left(\mathbf{r} - \mathbf{r}'  \right)$ the long-range part of the independent of the inner degrees of freedom (such as their electric and magnetic moments) potential energy of $i$-th and $j$-th interacting particles located in points  $\mathbf{r}$ and $\mathbf{r}'$, respectively.

The number of $i$-th kind particles in an infinitesimal volume $d\mathbf{r}$ near a point $\mathbf{r}$ is $n_i(\mathbf{r})\,d\mathbf{r}$. Similarly, the number of $j$-th kind particles in infinitesimal volume $d\mathbf{r}'$ near a point $\mathbf{r}'$ is $n_j(\mathbf{r}')\,d\mathbf{r}'$. Then the full energy of interactions between these particles near points $\mathbf{r}$ and $\mathbf{r}'$  is
\begin{equation}\label{U1}
n_i(\mathbf{r})\, d\mathbf{r}\ K_{ij}(\mathbf{r}-\mathbf{r}')\ n_j(\mathbf{r}')\,  d\mathbf{r}'.
\end{equation}
Hence the inner degrees of freedom independent part energy of  the system have a form:
\begin{equation}\label{Knn}
     U_1 = \frac{1}{2}\sum_{i,j=1}^{m}\,   \iint\limits_{(V)}
    K_{ij}(\mathbf{r}-\mathbf{r}')\, n_i(\mathbf{r})\,
    n_j(\mathbf{r}')\, d\mathbf{r}\, d\mathbf{r}'.
\end{equation}
The similar way leads to expression for configuration part of the free energy with account of the atomic electric and magnetic moments of the components:
\begin{equation}\label{config}
\begin{array}{r}
 {\displaystyle   U=\frac{1}{2}\sum_{i,j=1}^{m}\, \iint\limits_{(V)}\,
    K_{ij}(\mathbf{r}-\mathbf{r}')\, n_i(\mathbf{r})\,
    n_j(\mathbf{r}')\, d\mathbf{r}\, d\mathbf{r}' + } \\
{\displaystyle + \frac{1}{2}\sum_{i,j=1}^{m}\, \iint\limits_{(V)}\,
    Q_{ij}(\mathbf{r}-\mathbf{r}') n_i(\mathbf{r})
    n_j(\mathbf{r}') \left( \mathbf{D}_i(\mathbf{r})\cdot
    \mathbf{D}_j(\mathbf{r}') \right) d\mathbf{r} d\mathbf{r}'+ } \\
{\displaystyle + \frac{1}{2}\sum_{i,j=1}^{m}\, \iint\limits_{(V)}\,
    R_{ij}(\mathbf{r}-\mathbf{r}') n_i(\mathbf{r})
    n_j(\mathbf{r}') \left( \mathbf{M}_i(\mathbf{r}) \cdot
    \mathbf{M}_j(\mathbf{r}') \right) d\mathbf{r} d\mathbf{r}'+}\\
{\displaystyle + \frac{1}{2}\sum_{i,j=1}^{m}\, \iint\limits_{(V)}\,
    S_{ij}(\mathbf{r}-\mathbf{r}') n_i(\mathbf{r})
    n_j(\mathbf{r}') \left( \mathbf{D}_i(\mathbf{r}) \cdot
    \mathbf{M}_j(\mathbf{r}') \right) d\mathbf{r} d\mathbf{r}', }
\end{array}
\end{equation}
where $ \mathbf{D}_i(\mathbf{r})$ and $\mathbf{M}_i(\mathbf{r})$ are electric and magnetic atomic moments of $i$-th components, located at the point
$\mathbf{r}$,  $ K_{ij}(\mathbf{r}-\mathbf{r}') $, $ Q_{ij}(\mathbf{r}-\mathbf{r}') $, $ R_{ij}(\mathbf{r}-\mathbf{r}') $ and $ S_{ij}(\mathbf{r}-\mathbf{r}') $ are long-range parts of the relevant two-body interatomic potentials.

It should be noted that the vacancies do not contribute into the configuration free energy, but their contribution into the free energy due to entropy term:
\begin{equation}\label{entropy}
   S = - \sum_{i=0}^m \int\limits_{(V)} n_i(\mathbf{r}) \ln
    \left(\frac{n_i(\mathbf{r})}{n(\mathbf{r})}\right)
    d\mathbf{r},
\end{equation}
where ${\displaystyle n(\mathbf{r})}$ is the summarized local density of the particles and vacancies
\begin{equation}\label{n(r)}
    n(\mathbf{r})=\sum_{j=0}^m n_j(\mathbf{r}).
\end{equation}

Thus, the Helmholtz free energy $F$ of the system with account of the non-uniform external electric $\mathbf{E}(\mathbf{r})$ and magnetic $\mathbf{H} (\mathbf{r})$ fields has the following form:
\begin{equation}\label{helmholtz}
\begin{array}{r}
 {\displaystyle   F = \frac{1}{2}\sum_{i,j=1}^{m}\, \iint\limits_{(V)}
    K_{ij}(\mathbf{r}-\mathbf{r}') n_i(\mathbf{r})
    n_j(\mathbf{r}') d\mathbf{r}\, d\mathbf{r}' + } \\
{\displaystyle + \frac{1}{2}\sum_{i,j=1}^{m}\, \iint\limits_{(V)}
    Q_{ij}(\mathbf{r}-\mathbf{r}') n_i(\mathbf{r})
    n_j(\mathbf{r}') \left( \mathbf{D}_i(\mathbf{r})\cdot
    \mathbf{D}_j(\mathbf{r}') \right) d\mathbf{r}\, d\mathbf{r}'+ } \\
{\displaystyle + \frac{1}{2}\sum_{i,j=1}^{m}\, \iint\limits_{(V)}
    R_{ij}(\mathbf{r}-\mathbf{r}') n_i(\mathbf{r})
    n_j(\mathbf{r}') \left( \mathbf{M}_i(\mathbf{r})\cdot
    \mathbf{M}_j(\mathbf{r}') \right) d\mathbf{r}\, d\mathbf{r}'+ }\\
{\displaystyle + \frac{1}{2}\sum_{i,j=1}^{m}\, \iint\limits_{(V)}
    S_{ij}(\mathbf{r}-\mathbf{r}') n_i(\mathbf{r})
    n_j(\mathbf{r}') \left( \mathbf{D}_i(\mathbf{r}) \cdot
    \mathbf{M}_j(\mathbf{r}') \right) d\mathbf{r} d\mathbf{r}' + }\\
{\displaystyle + \sum_{i=1}^m\, \int\limits_{(V)} \left( \mathbf{E}(\mathbf{r}) \cdot
\mathbf{D}_i(\mathbf{r}) \right) n_i(\mathbf{r})\,  d\mathbf{r} +
\sum_{i=1}^m\, \int\limits_{(V)} \left( \mathbf{H}(\mathbf{r}) \cdot
\mathbf{M}_i(\mathbf{r}) \right) n_i(\mathbf{r})\,  d\mathbf{r} +}\\
{\displaystyle   + T\sum_{i=0}^m\, \int\limits_{(V)} n_i(\mathbf{r}) \ln
    \left(\frac{n_i(\mathbf{r})}{n(\mathbf{r})}\right)
    d\mathbf{r}},
\end{array}
\end{equation}
where $T$ is absolute temperature in energetic units.

The equilibrium space distributions of components is determined by minimum of Helmholtz free energy functional with account of the  conditions~(\ref{packing}) and~(\ref{N}). But it should be noted that beyond this condition we need some additional conditions connecting all the electric $\mathbf{D}_i(\mathbf{r})$ and magnetic $\mathbf{M}_i(\mathbf{r})$ moments with the external fields. Suppose that external fields influence on the orientations of related moments but do not influence on their magnitudes. Hence we have
\begin{equation}\label{D-M}
    \left(\mathbf{D}_i\left(\mathbf{r}  \right)\right)^2 - D_i^2 = 0; \quad     \left(\mathbf{M}_i\left(\mathbf{r}  \right)\right)^2 - M_i^2 = 0.
\end{equation}
\subsection{The Lagrange functional and full equilibrium in the system}

For the minimization of the Helmholtz free energy~(\ref{helmholtz}) at the conditions~(\ref{packing}), (\ref{N}), (\ref{D-M}) let us introduce the Lagrange functional $\mathcal{L}$ depending on the local densities $n_i(\mathbf{r})$ of the components, their electric and magnetic moments $\mathbf{D}_i(\mathbf{r})$, $\mathbf{M}_i(\mathbf{r})$, the external fields $\mathbf{E}(\mathbf{r})$, $\mathbf{H}(\mathbf{r})$, and the Lagrange multipliers $\Psi(\mathbf{r})$, $\lambda_i(\mathbf{r})$, $\nu_i(\mathbf{r})$, $\mu_i$:
\begin{equation}\label{lagrange}
\begin{array}{r}
{\displaystyle
    \mathcal{L}(\{n_i(\mathbf{r})\},\, \left\{ \mathbf{D}_i\left(\mathbf{r}  \right) \right\},\, \left\{ \mathbf{M}_i\left(\mathbf{r}  \right)  \right\}, \,   \{\lambda_i\left( \mathbf{r} \right)\},\, \{\nu_i\left( \mathbf{r} \right)\},\, \{\Psi(\mathbf{r})\},\,
    \mu_i) = } \\ \\
{\displaystyle = F - \sum_{i=0}^{m}\mu_i\left[ \int\limits_{(V)}\
    n_i(\mathbf{r})\ d\mathbf{r}-N_i \right]
- \sum_{i=1}^m\ \int\limits_{(V)} \frac{\lambda_i\left( \mathbf{r} \right)}{2} \left[ \left(\mathbf{D}_i\left(\mathbf{r}  \right)\right)^2 - D_i^2 \right]\, d\mathbf{r}\ -
}\\ \\
{\displaystyle -
\sum_{i=1}^m\ \int\limits_{(V)} \frac{\nu_i\left( \mathbf{r} \right)}{2} \left[ \left(\mathbf{M}_i\left(\mathbf{r}  \right)\right)^2 - M_i^2 \right]\, d\mathbf{r}  - \int\limits_{(V)}\Psi(\mathbf{r}) \left(
\sum_{i=0}^m\ \omega_i\
    n_i(\mathbf{r})-1 \right) d\mathbf{r}.}
\end{array}
\end{equation}

The necessary condition or the extremum of the Lagrange functional is vanishing of the ordinary and functional derivatives with respect to $n_i(\mathbf{r})$, $\mathbf{D}_i(\mathbf{r})$, $\mathbf{M}_i(\mathbf{r})$, $\Psi(\mathbf{r})$, $\lambda_i(\mathbf{r})$, $\nu_i(\mathbf{r})$, $\mu_i$:
\begin{equation}\label{dLd-var}
    \left\{
\begin{array}{l}
{\displaystyle  \frac{\delta \mathcal{L}}{\delta n_i\left( \mathbf{r} \right)} = 0;}\\
{\displaystyle  \frac{\delta \mathcal{L}}{\delta \mathbf{D}_i\left( \mathbf{r} \right)} = 0;}\\
{\displaystyle  \frac{\delta \mathcal{L}}{\delta \mathbf{M}_i\left( \mathbf{r} \right)} = 0;}\\
{\displaystyle  \frac{\delta \mathcal{L}}{\delta \Psi_i\left( \mathbf{r} \right)} = 0;}\\
{\displaystyle  \frac{\partial \mathcal{L}}{\partial \mu_i\left( \mathbf{r} \right)} = 0;}\\
{\displaystyle  \frac{\delta \mathcal{L}}{\delta \lambda_i\left( \mathbf{r} \right)} = 0;}\\
{\displaystyle  \frac{\delta \mathcal{L}}{\delta \nu_i\left( \mathbf{r} \right)} = 0.}\
\end{array}
  \right.
\end{equation}

Simple calculations lead to following system of equations:
\begin{equation}\label{ext-n}
\begin{array}{rcl}
 & & {\displaystyle \frac{\delta \mathcal{L}}{\delta n_i\left( \mathbf{r} \right)} = 0 \Longrightarrow }\\
{\displaystyle \mu_i } & = & {\displaystyle - \omega_i \Psi(\mathbf{r}) + T \ln \left(\frac{n_i(\mathbf{r})}{n(\mathbf{r})}\right) + \sum_{j=1}^m \ \int\limits_{(V)} K_{ij}(\mathbf{r}-\mathbf{r}') n_j(\mathbf{r}') d\mathbf{r}' + }\\
& & {\displaystyle \quad   +  \sum_{j=1}^m\ \int\limits_{(V)} Q_{ij}(\mathbf{r}-\mathbf{r}')  \left(\mathbf{D}_i (\mathbf{r}) \cdot \mathbf{D}_j (\mathbf{r}')  \right) n_j(\mathbf{r}') d\mathbf{r}' +}\\
& & {\displaystyle \quad  + \sum_{j=1}^m\ \int\limits_{(V)} R_{ij}(\mathbf{r}-\mathbf{r}')  \left(\mathbf{M}_i (\mathbf{r}) \cdot \mathbf{M}_j (\mathbf{r}')  \right) n_j(\mathbf{r}') d\mathbf{r}' +}\\
& & {\displaystyle \quad  + \sum_{j=1}^m\ \int\limits_{(V)} S_{ij}(\mathbf{r}-\mathbf{r}') \frac{ \left\{ \left(\mathbf{D}_i (\mathbf{r}) \cdot \mathbf{M}_j (\mathbf{r}')  \right) + \left(\mathbf{M}_i (\mathbf{r}) \cdot \mathbf{D}_j (\mathbf{r}')  \right) \right\} }{2}  n_j(\mathbf{r}') d\mathbf{r}' + }\\
& & {\displaystyle \quad + \left( \mathbf{E}\left( \mathbf{r} \right) \cdot \mathbf{D}_i\left( \mathbf{r} \right) \right) + \left( \mathbf{H}\left( \mathbf{r} \right) \cdot \mathbf{M}_i\left( \mathbf{r} \right) \right); }
\end{array}
\end{equation}

\begin{equation}\label{ext-D}
\begin{array}{l}
{\displaystyle  \frac{\delta \mathcal{L}}{\delta \mathbf{D}_i\left( \mathbf{r} \right)} = 0 \Longrightarrow}\\
{\displaystyle \sum_{j=1}^m\ \int\limits_{(V)} n_i(\mathbf{r}) \, Q_{ij}\left( \mathbf{r} - \mathbf{r}' \right) n_j(\mathbf{r}')\,\mathbf{D}_j(\mathbf{r}')\, d\mathbf{r}' + }\\
{\displaystyle \qquad  + \frac{1}{2}\, \sum_{j=1}^m\ \int\limits_{(V)} n_i(\mathbf{r}) \, S_{ij}\left( \mathbf{r} - \mathbf{r}' \right) n_j(\mathbf{r}')\,\mathbf{M}_j(\mathbf{r}')\, d\mathbf{r}'  - \lambda_{i}(\mathbf{r}) \mathbf{D}_i (\mathbf{r}) + \mathbf{E}(\mathbf{r}) n_i(\mathbf{r}) = 0;}
\end{array}
\end{equation}

\begin{equation}\label{ext-M}
\begin{array}{l}
{\displaystyle  \frac{\delta \mathcal{L}}{\delta \mathbf{M}_i\left( \mathbf{r} \right)} = 0\ \Longrightarrow }\\
{\displaystyle \sum_{j=1}^m\ \int\limits_{(V)} n_i(\mathbf{r}) \, R_{ij}\left( \mathbf{r} - \mathbf{r}' \right) n_j(\mathbf{r}')\,\mathbf{M}_j(\mathbf{r}')\, d\mathbf{r}' + }\\
{\displaystyle \qquad  + \frac{1}{2}\, \sum_{j=1}^m\ \int\limits_{(V)} n_i(\mathbf{r}) \, S_{ij}\left( \mathbf{r} - \mathbf{r}' \right) n_j(\mathbf{r}')\,\mathbf{D}_j(\mathbf{r}')\, d\mathbf{r}'  - \nu_{i}(\mathbf{r}) \mathbf{M}_i (\mathbf{r}) + \mathbf{H}(\mathbf{r}) n_i(\mathbf{r}) = 0;}
\end{array}
\end{equation}
\begin{equation}\label{ext-Psi}
\begin{array}{r}
    {\displaystyle \frac{\delta \mathcal{L}}{\delta \Psi_i\left( \mathbf{r} \right)} = 0 \Longrightarrow  }\\
{\displaystyle  \sum_{i=0}^m\ \omega_i\ n_i(\mathbf{r})-1=0;}
\end{array}
\end{equation}

\begin{equation}\label{ext-mu}
\begin{array}{r}
{\displaystyle \frac{\partial \mathcal{L}}{\partial \mu_i\left( \mathbf{r} \right)} = 0 \Longrightarrow}\\
{\displaystyle \int\limits_{(V)}\ n_i(\mathbf{r})\ d\mathbf{r}  - N_i=0;}
\end{array}
\end{equation}

\begin{equation}\label{ext-lambda}
\begin{array}{r}
    {\displaystyle  \frac{\delta \mathcal{L}}{\delta \lambda_i\left( \mathbf{r} \right)} = 0 \Longrightarrow }\\
{\displaystyle \left(\mathbf{D}_i\left(\mathbf{r}  \right)\right)^2 - D_i^2 = 0;}
\end{array}
\end{equation}

\begin{equation}\label{ext-nu}
\begin{array}{r}
{\displaystyle \frac{\delta \mathcal{L}}{\delta \nu_i\left( \mathbf{r} \right)} = 0 \Longrightarrow }\\
{\displaystyle \left(\mathbf{M}_i\left(\mathbf{r}  \right)\right)^2 - M_i^2 = 0. }
\end{array}
\end{equation}

This system of integral equations~(\ref{ext-n}-\ref{ext-nu}) describes the space distributions of the components as well electric and magnetic moments in the system with account of both short-range and  long-range parts on interatomic potentials in presence of the external fields. This system of equations take place for the case of full thermodynamic equilibrium in the systems.

Unfortunately, at present there are no effective methods of such kind equations solutions for kernel $K_{ij}(\mathbf{r})$, $Q_{ij}(\mathbf{r})$, $R_{ij}(\mathbf{r})$, $S_{ij}(\mathbf{r})$ of general form.
But under the some conditions this system of equations can be reduced to a system of the partial differential equations. Instead of the interatomic potentials $K_{ij}(\mathbf{r})$, $Q_{ij}(\mathbf{r})$, $R_{ij}(\mathbf{r})$, $S_{ij}(\mathbf{r})$, the system of partial differential equations contains a set of integral characteristics of the potentials.

\section{Ginzburg-Landau-Cahn-Hilliard-like approximation}

\subsection{Reduction of the Generalized lattice model to Ginzburg-Landau-Cahn-Hilliard-like approximation}

There are at least three scales of the sizes in the system:
\begin{enumerate}
    \item atomic sizes $a_0$;
    \item range of actions of long-range parts of the interatomic potentials $r_0$;
    \item distances $b_0$ on which changes local compositions and/or local moments of the components in the system.
\end{enumerate}
Suppose these parameters obey the inequalities:
\begin{equation}\label{arb}
    a_0\lesssim r_0 \ll b_0.
\end{equation}

Let us transform the expression for moments-independent part of the configuration energy in
~(\ref{helmholtz}):
\begin{equation}\label{F-1}
F_{1}=\frac{1}{2}\,\sum_{i,j=1}^{m}\ \iint\limits_{(V)}
    K_{ij}(\mathbf{r}-\mathbf{r}')\, n_i(\mathbf{r})
    n_j(\mathbf{r}')\, d\mathbf{r}\, d\mathbf{r}'.
\end{equation}
With regard to condition
\begin{equation}\label{r-r'}
    \mid \mathbf{r}-\mathbf{r}' \mid \ll b_0
\end{equation}
we have
\begin{equation}\label{nj}
    n_j(\mathbf{r}') \approx n_j(\mathbf{r}) + \sum_{s=1}^3 \frac{\partial
    n_j(\mathbf{r})}{\partial x_s} (x_s'- x_s)
    + \frac{1}{2} \sum_{s_1,s_2=1}^3 \frac{\partial^{\,2} n_j
    (\mathbf{r})}{\partial x_{s_1}\, \partial x_{s_2}}
    (x_{s_1}'-x_{s_1})(x_{s_2}'-x_{s_2}),
\end{equation}
where $x_s$ and $x'_s$ denote Cartesian components of the vectors $\mathbf{r} $ and  $\mathbf{r}'$, respectively.
Substitution of~(\ref{nj}) into~(\ref{F-1}) lead to the result:
\begin{equation}\label{Fn}
    F_1 \left( \left\{n_k(\mathbf{r}) \right\} \right) = \int\limits_{(V)} \,
\left\{\frac{1}{12} \sum_{i,j=1}^m
K_{ij}^{(2)} \,n_i(\mathbf{r})\, \Delta
n_j(\mathbf{r}) + \frac{1}{2} \sum_{i,j=1}^m K_{ij}^{(0)}
\, n_i(\mathbf{r})\,  n_j(\mathbf{r})
  \right\}\, d\mathbf{r},
\end{equation}
where $\Delta$ is the Laplace operator, $K_{ij}^{(p)}$ are some integral characteristics of the interatomic potentials
\begin{equation}\label{Kij}
    K_{ij}^{(p)} = \int\limits_{(V)} \, K_{ij}(\mathbf{r}')|\mathbf{r}'|^{p}\,
    d \mathbf{r}'.
\end{equation}

The expressions for moment-depending parts of configuration energy in~(\ref{helmholtz}) can be transformed by analogy:
\begin{equation}\label{FD}
\begin{array}{r}
{\displaystyle  F_2( \left\{n_k(\mathbf{r}) \right\}, \left\{\mathbf{D}_k(\mathbf{r}) \right\}  ) = \int\limits_{(V)} \, \left\{\frac{1}{12} \sum_{i,j=1}^m
Q_{ij}^{(2)} \, \left( \mathbf{D}_i(\mathbf{r}) n_i(\mathbf{r})\cdot  \Delta \left[
\mathbf{D}_j (\mathbf{r}) n_j(\mathbf{r}) \right] \right) \right. + }\\
{\displaystyle \left. + \frac{1}{2} \sum_{i,j=1}^m Q_{ij}^{(0)}
\, \left(  \mathbf{D}_i(\mathbf{r}) \cdot \mathbf{D}_j(\mathbf{r})
\right) n_i(\mathbf{r})\,n_j(\mathbf{r})
  \right\}\, d\mathbf{r},}
\end{array}
\end{equation}

\begin{equation}\label{FM}
\begin{array}{r}
{\displaystyle  F_3( \left\{n_k(\mathbf{r}) \right\}, \left\{\mathbf{M}_k(\mathbf{r}) \right\}  ) = \int\limits_{(V)} \, \left\{\frac{1}{12} \sum_{i,j=1}^m
R_{ij}^{(2)} \, \left( \mathbf{M}_i(\mathbf{r}) n_i(\mathbf{r})\cdot  \Delta \left[
\mathbf{M}_j (\mathbf{r}) n_j(\mathbf{r}) \right] \right) \right. + }\\
{\displaystyle \left. + \frac{1}{2} \sum_{i,j=1}^m R_{ij}^{(0)}
\, \left(  \mathbf{M}_i(\mathbf{r}) \cdot \mathbf{M}_j(\mathbf{r})
\right) n_i(\mathbf{r})\,n_j(\mathbf{r})
  \right\}\, d\mathbf{r},}
\end{array}
\end{equation}

\begin{equation}\label{FDM}
\begin{array}{r}
{\displaystyle   F_4( \left\{n_k(\mathbf{r}) \right\}, \left\{\mathbf{D}_k(\mathbf{r}) \right\}, \left\{\mathbf{M}_k(\mathbf{r}) \right\}   ) = \int\limits_{(V)} \, \left\{\frac{1}{12} \sum_{i,j=1}^m
S_{ij}^{(2)} \, \left( \mathbf{D}_i(\mathbf{r}) n_i(\mathbf{r})\cdot  \Delta \left[
\mathbf{M}_j (\mathbf{r}) n_j(\mathbf{r}) \right] \right) \right. + }\\
{\displaystyle \left. + \frac{1}{2} \sum_{i,j=1}^m S_{ij}^{(0)}
\, \left(  \mathbf{D}_i(\mathbf{r}) \cdot \mathbf{M}_j(\mathbf{r})
\right) n_i(\mathbf{r})\,n_j(\mathbf{r})
  \right\}\, d\mathbf{r},}
\end{array}
\end{equation}
where $Q_{ij}^{(p)} $, $R_{ij}^{(p)} $ and $S_{ij}^{(p)} $ are the integral  characteristics of the moment-depending parts of the interatomic potentials determined by perfect analogy with~(\ref{Kij}).

Substitute (\ref{Kij}), (\ref{FD}), (\ref{FM}) and (\ref{FDM}) into  (\ref{helmholtz}):
\begin{equation}\label{FGLCH}
\begin{array}{r}
{\displaystyle  F\left( \left\{n_k(\mathbf{r}) \right\}, \left\{\mathbf{D}_k(\mathbf{r}) \right\},
 \left\{\mathbf{M}_k(\mathbf{r}) \right\}  \right) = \int\limits_{(V)}
\biggl\{  \frac{1}{12} \sum_{i,j=1}^m \biggl[
K_{ij}^{(2)} \,n_i(\mathbf{r})\, \Delta
n_j(\mathbf{r}) +   } \\ \\
{\displaystyle +  Q_{ij}^{(2)} \, \left( \mathbf{D}_i(\mathbf{r}) n_i(\mathbf{r})\cdot  \Delta \left[
\mathbf{D}_j (\mathbf{r}) n_j(\mathbf{r}) \right] \right)  + R_{ij}^{(2)} \,
\left( \mathbf{M}_i(\mathbf{r}) n_i(\mathbf{r})\cdot  \Delta \left[
\mathbf{M}_j (\mathbf{r}) n_j(\mathbf{r}) \right] \right) + }\\ \\
{\displaystyle + S_{ij}^{(2)} \,
\left( \mathbf{D}_i(\mathbf{r}) n_i(\mathbf{r})\cdot  \Delta \left[
\mathbf{M}_j (\mathbf{r}) n_j(\mathbf{r}) \right] \right)\biggr] + }\\ \\
{\displaystyle + \frac{1}{2} \sum_{i,j=1}^m \biggl[ K_{ij}^{(0)}
 + Q_{ij}^{(0)} \left(  \mathbf{D}_i(\mathbf{r})\cdot \mathbf{D}_j(\mathbf{r})
 \right) + R_{ij}^{(0)} \left(  \mathbf{M}_i(\mathbf{r}) \cdot \mathbf{M}_j(\mathbf{r})
 \right) + } \\ \\
{\displaystyle + S_{ij}^{(0)} \left(  \mathbf{D}_i(\mathbf{r}) \cdot \mathbf{M}_j(\mathbf{r})
 \right) \biggr] \, n_i(\mathbf{r})\,  n_j(\mathbf{r}) +  }\\ \\
{\displaystyle + \sum_{i=1}^m \biggl[ \left( \mathbf{E}(\mathbf{r}) \cdot
\mathbf{D}_i(\mathbf{r}) \right) n_i(\mathbf{r}) + \left( \mathbf{H}(\mathbf{r}) \cdot
\mathbf{M}_i(\mathbf{r}) \right) n_i(\mathbf{r}) + T\, n_i(\mathbf{r}) \ln
\left(\frac{n_i(\mathbf{r})}{n(\mathbf{r})}\right) \biggr] }\biggr\} \, d\mathbf{r}.
\end{array}
\end{equation}

The next step is using the Green formula
\begin{equation}\label{Green}
    \int\limits_{(V)}\, u\left( \mathbf{r} \right) \Delta v\left( \mathbf{r} \right)\,d\mathbf{r} = - \int\limits_{(V)}\, \left(  \nabla u\left( \mathbf{r} \right)\cdot \nabla v\left( \mathbf{r} \right) \right) d\mathbf{r},
\end{equation}
that has place if the functions $u\left( \mathbf{r} \right)$, $v\left( \mathbf{r} \right)$ and their gradients $\nabla u\left( \mathbf{r} \right)$, $\nabla v\left( \mathbf{r} \right)$ vanish on the boundary of domain~$V$.
Using this formula, the terms with Laplacians in~(\ref{FGLCH}) take the following forms:
\begin{equation}\label{K-2}
    \int\limits_{(V)} \, n_i\left( \mathbf{r} \right)\,\Delta n_j\left( \mathbf{r} \right)\,d\mathbf{r} = - \int\limits_{(V)} \,\left(\nabla n_i\left( \mathbf{r} \right)\cdot \nabla n_j\left( \mathbf{r} \right)  \right) d\mathbf{r},
\end{equation}

\begin{equation}\label{Q-2}
    \int\limits_{(V)} \, \left( \mathbf{D}_i \left( \mathbf{r} \right) n_i\left( \mathbf{r} \right)\cdot \Delta \left[ \mathbf{D}_j \left( \mathbf{r} \right) n_j\left( \mathbf{r} \right)  \right]  \right)\,d\mathbf{r} = - \int\limits_{(V)} \sum_{\alpha=1}^3 \left(\nabla \left[ D_i^{\alpha} (\mathbf{r})  n_i\left( \mathbf{r} \right) \right] \cdot \nabla \left[  D_j^{\alpha}  (\mathbf{r}) n_j\left( \mathbf{r} \right) \right] \right) d\mathbf{r},
\end{equation}

\begin{equation}\label{R-2}
    \int\limits_{(V)} \, \left( \mathbf{M}_i \left( \mathbf{r} \right) n_i\left( \mathbf{r} \right)\cdot \Delta \left[ \mathbf{M}_j \left( \mathbf{r} \right) n_j\left( \mathbf{r} \right)  \right]  \right)\,d\mathbf{r} = - \int\limits_{(V)} \sum_{\alpha=1}^3 \left(\nabla \left[ M_i^{\alpha} (\mathbf{r})  n_i\left( \mathbf{r} \right) \right] \cdot \nabla \left[  M_j^{\alpha}  (\mathbf{r}) n_j\left( \mathbf{r} \right) \right] \right) d\mathbf{r},
\end{equation}

\begin{equation}\label{S-2}
    \int\limits_{(V)} \, \left( \mathbf{D}_i \left( \mathbf{r} \right) n_i\left( \mathbf{r} \right)\cdot \Delta \left[ \mathbf{M}_j \left( \mathbf{r} \right) n_j\left( \mathbf{r} \right)  \right]  \right)\,d\mathbf{r} = - \int\limits_{(V)} \sum_{\alpha=1}^3 \left(\nabla \left[ D_i^{\alpha} (\mathbf{r})  n_i\left( \mathbf{r} \right) \right] \cdot \nabla \left[  M_j^{\alpha}  (\mathbf{r}) n_j\left( \mathbf{r} \right) \right] \right) d\mathbf{r},
\end{equation}
where $ D_i^{\alpha} (\mathbf{r}) $ and $ M_i^{\alpha} (\mathbf{r}) $ denote components of the vectors $\mathbf{D}_i \left( \mathbf{r} \right)$ and $\mathbf{M}_i \left( \mathbf{r} \right)$, respectively.
As a result, we have the following expression for the Helmholtz free energy functional:
\begin{equation}\label{FGLCH2}
\begin{array}{r}
{ \displaystyle  F\left( \left\{n_k(\mathbf{r}) \right\}, \left\{\mathbf{D}_k(\mathbf{r}) \right\},
 \left\{\mathbf{M}_k(\mathbf{r}) \right\}  \right) = \int\limits_{(V)}
\biggl\{ - \frac{1}{12} \sum_{i,j=1}^m \biggl[
K_{ij}^{(2)} \left(\nabla n_i(\mathbf{r})\cdot \nabla
n_j(\mathbf{r}) \right) +   } \\ \\
{\displaystyle +  Q_{ij}^{(2)} \sum_{\alpha=1}^3 \left( \nabla \left[ {D}^{\alpha}_i(\mathbf{r}) n_i(\mathbf{r}) \right] \cdot  \nabla \left[
{D}^{\alpha}_j (\mathbf{r}) n_j(\mathbf{r}) \right] \right)  +   } \\ \\
{\displaystyle + R_{ij}^{(2)} \,
\sum_{\alpha=1}^3 \left( \nabla \left[ {M}^{\alpha}_i(\mathbf{r}) n_i(\mathbf{r}) \right] \cdot  \nabla \left[
{M}^{\alpha}_j (\mathbf{r}) n_j(\mathbf{r}) \right] \right) + }\\ \\
{\displaystyle + S_{ij}^{(2)} \,
\sum_{\alpha=1}^3 \left( \nabla \left[ {D}^{\alpha}_i(\mathbf{r}) n_i(\mathbf{r}) \right] \cdot  \nabla \left[
{M}^{\alpha}_j (\mathbf{r}) n_j(\mathbf{r}) \right] \right)\biggr] + }\\ \\
{\displaystyle + \frac{1}{2} \sum_{i,j=1}^m \biggl[ K_{ij}^{(0)}
 + Q_{ij}^{(0)} \left(  \mathbf{D}_i(\mathbf{r})\cdot \mathbf{D}_j(\mathbf{r})
 \right) + R_{ij}^{(0)} \left(  \mathbf{M}_i(\mathbf{r}) \cdot \mathbf{M}_j(\mathbf{r})
 \right)
+ } \\ \\
{\displaystyle
+ S_{ij}^{(0)} \left(  \mathbf{D}_i(\mathbf{r}) \cdot \mathbf{M}_j(\mathbf{r})
 \right) \biggr] \, n_i(\mathbf{r})\,  n_j(\mathbf{r}) +  }\\ \\
{\displaystyle + \sum_{i=1}^m \biggl[ \left( \mathbf{E}(\mathbf{r}) \cdot
\mathbf{D}_i(\mathbf{r}) \right) n_i(\mathbf{r}) + \left( \mathbf{H}(\mathbf{r}) \cdot
\mathbf{M}_i(\mathbf{r}) \right) n_i(\mathbf{r}) + T\, n_i(\mathbf{r}) \ln
\left(\frac{n_i(\mathbf{r})}{n(\mathbf{r})}\right) \biggr] }\biggr\} \, d\mathbf{r}.
\end{array}
\end{equation}

This functional is similar to Ginzburg-Landau and Cahn-Hilliard functionals, but in contrast to these functionals formula~(\ref{FGLCH2})
\begin{enumerate}
    \item is not restricted by the polynomial over order parameters in the integrand;
    \item all the parameters in~(\ref{FGLCH2}) have clear physical sense.
\end{enumerate}

The functional~(\ref{FGLCH2}) has some essential advantages over
the more exact functional~(\ref{helmholtz}):
\begin{itemize}
    \item The expression~(\ref{FGLCH2}) does not contain any unknown
    functions of general form (such as interatomic potentials), but contains
    the finite set of their simple characteristics (numerical parameters) $K_{ij}^{(p)}$ (\ref{Kij});
    \item Analysis and solution of differential equations, that can
    be obtained from functional (\ref{FGLCH2}), are much more
    simple problems as solution of non-linear integral
    equations~(\ref{ext-n}--\ref{ext-nu}) for general case;
    \item The inverse problem~--- search of parameters using some
    experimental data (such as phase diagrams)~--- is the immense
    problem from the integral equations~(\ref{ext-n}--\ref{ext-nu}),
    but quite realistic one from the functional (\ref{FGLCH2});
    \item At last, the solution, based on the
    functional~(\ref{FGLCH2}), can be used as well initial
    approximation for much more complicated problems,
    related to functional (~(\ref{ext-n}--\ref{ext-nu})).
\end{itemize}

There is the necessary condition of the mathematical correctness for the functional~(\ref{FGLCH2}). The necessity of the fluctuations suppression in the system means that all the matrices $K_{ij}^{(2)}$, $Q_{ij}^{(2)}$, $R_{ij}^{(2)}$, $S_{ij}^{(2)}$ must be negative-definite.

\subsection{The Lagrange functional for GLCH-like approximation and full equilibrium in the system}

It should be noted the conditions~(\ref{packing}), (\ref{N}), (\ref{D-M}) must satisfied not only for functional~(\ref{helmholtz}), but also for functional~(\ref{FGLCH2}).
Let us introduce the Lagrange functional $\mathcal{L}_1$, corresponding to the GLCH-like Helmholtz free energy functional~(\ref{FGLCH2}), by analogy with~(\ref{lagrange}):
\begin{equation}\label{lagr2}
\begin{array}{r}
{\displaystyle
    \mathcal{L}_1 (\{n_k(\mathbf{r})\}, \left\{\mathbf{M}_k(\mathbf{r}) \right\}, \left\{\mathbf{D}_k(\mathbf{r}) \right\}, \{\lambda_k\left( \mathbf{r} \right)\},\, \{\nu_k\left( \mathbf{r} \right)\},\, \{\Psi(\mathbf{r})\},\,
    \mu_k) =  } \\ \\
{\displaystyle = F\left( \left\{n_k(\mathbf{r}) \right\}, \left\{\mathbf{D}_k(\mathbf{r}) \right\},
 \left\{\mathbf{M}_k(\mathbf{r}) \right\}  \right) - } \\ \\
{\displaystyle  - \sum_{i=0}^{m}\mu_i\left[ \int\limits_{(V)}\
    n_i(\mathbf{r})\ d\mathbf{r}-N_i \right]
- \sum_{i=1}^m\ \int\limits_{(V)} \frac{\lambda_i\left( \mathbf{r} \right)}{2} \left[ \left(\mathbf{D}_i\left(\mathbf{r}  \right)\right)^2 - D_i^2 \right]\, d\mathbf{r}\ -
}\\ \\
{\displaystyle -
\sum_{i=1}^m\ \int\limits_{(V)} \frac{\nu_i\left( \mathbf{r} \right)}{2} \left[ \left(\mathbf{M}_i\left(\mathbf{r}  \right)\right)^2 - M_i^2 \right]\, d\mathbf{r}  - \int\limits_{(V)}\Psi(\mathbf{r}) \left(
\sum_{i=0}^m\ \omega_i\
    n_i(\mathbf{r})-1 \right) d\mathbf{r}. }
\end{array}
\end{equation}

This functional has the following form:
\begin{equation}\label{lagr3}
\begin{array}{r}
     {\displaystyle \mathcal{L}_1 (\{n_k(\mathbf{r})\}, \left\{\mathbf{M}_k(\mathbf{r}) \right\}, \left\{\mathbf{D}_k(\mathbf{r}) \right\}, \{\lambda_k\left( \mathbf{r} \right)\},\, \{\nu_k\left( \mathbf{r} \right)\},\, \{\Psi(\mathbf{r})\},\,
    \mu_k) =}\\  \\
{\displaystyle
= \widetilde{\mathcal{L}_1}+ \widetilde{\widetilde{\mathcal{L}_1}},}
\end{array}
   \end{equation}
where
\begin{equation}\label{L1}
     \widetilde{\mathcal{L}_1} = \sum_{i=1}^m \left[ \mu_i N_i + \frac{D_i^2}{2} \int\limits_{(V)} \lambda_i(\mathbf{r})\, d\mathbf{r} + \frac{M_i^2}{2} \int\limits_{(V)} \nu_i(\mathbf{r})\, d\mathbf{r} \right] + \int\limits_{(V)} \Psi(\mathbf{r}) d\mathbf{r}
\end{equation}
is the part of the functional~(\ref{lagr2}), which does not depend in explicit form on functions $\mathbf{D}_i(\mathbf{r})$, $\mathbf{M}_i(\mathbf{r})$, ${n}_i(\mathbf{r})$,
\begin{equation}\label{L2}
\begin{array}{r}
{\displaystyle     \widetilde{\widetilde{\mathcal{L}_1}} \left( \left\{n_k(\mathbf{r}) \right\}, \left\{\mathbf{D}_k(\mathbf{r}) \right\},
 \left\{\mathbf{M}_k(\mathbf{r}) \right\}  \right) = F\left( \left\{n_k(\mathbf{r}) \right\}, \left\{\mathbf{D}_k(\mathbf{r}) \right\},
 \left\{\mathbf{M}_k(\mathbf{r}) \right\}  \right) -}\\  \\
{\displaystyle - \int\limits_{(V)} \left\{ \sum_{i=1}^m \left[\mu_i\, n_i (\mathbf{r}) + \frac{\lambda_i(\mathbf{r})}{2}\, \mathbf{D}_i^2(\mathbf{r}) +  \frac{\nu_i(\mathbf{r})}{2}\, \mathbf{M}_i^2(\mathbf{r}) + \Psi(\mathbf{r})\,\omega_i\, n_i(\mathbf{r}) \right] \right\}\,d\mathbf{r}}
\end{array}
\end{equation}
is the second part of the functional~(\ref{lagr2}),
which depends on $n_i(\mathbf{r})$,  ${D}_i^{\alpha}(\mathbf{r})$, ${M}_i^{\alpha}(\mathbf{r})$ in explicit form.
The last functional has a form
\begin{equation}\label{L3}
\begin{array}{r}
 {\displaystyle  \widetilde{\widetilde{\mathcal{L}_1}}\left(  \left\{u_s(\mathbf{r})  \right\} \right) =  \int\limits_{(V)} \Lambda\left( u_s(\mathbf{r}),\, \nabla u_s(\mathbf{r})  \right)\, d\mathbf{r}},
\end{array}
\end{equation}
where $u_s(\mathbf{r})$ denotes all the functions $n_i(\mathbf{r})$,  ${D}_i^{\alpha}(\mathbf{r})$, ${M}_i^{\alpha}(\mathbf{r})$  and
\begin{equation}\label{Lambda}
\begin{array}{r}
{\displaystyle \Lambda\left( u_s(\mathbf{r}),\, \nabla u_s(\mathbf{r})  \right) =  - \frac{1}{12} \sum_{i,j=1}^m \biggl[
K_{ij}^{(2)} \left(\nabla n_i(\mathbf{r})\cdot \nabla
n_j(\mathbf{r}) \right) +   } \\ \\
{\displaystyle +  Q_{ij}^{(2)} \sum_{\alpha=1}^3 \left( \nabla \left[ {D}^{\alpha}_i(\mathbf{r}) n_i(\mathbf{r}) \right] \cdot  \nabla \left[
{D}^{\alpha}_j (\mathbf{r}) n_j(\mathbf{r}) \right] \right)  +   } \\ \\
{\displaystyle + R_{ij}^{(2)} \,
\sum_{\alpha=1}^3 \left( \nabla \left[ {M}^{\alpha}_i(\mathbf{r}) n_i(\mathbf{r}) \right] \cdot  \nabla \left[
{M}^{\alpha}_j (\mathbf{r}) n_j(\mathbf{r}) \right] \right) + }\\ \\
{\displaystyle + S_{ij}^{(2)} \,
\sum_{\alpha=1}^3 \left( \nabla \left[ {D}^{\alpha}_i(\mathbf{r}) n_i(\mathbf{r}) \right] \cdot  \nabla \left[
{M}^{\alpha}_j (\mathbf{r}) n_j(\mathbf{r}) \right] \right)\biggr] + }\\ \\
{\displaystyle + \frac{1}{2} \sum_{i,j=1}^m \biggl[ K_{ij}^{(0)}
 + Q_{ij}^{(0)} \left(  \mathbf{D}_i(\mathbf{r})\cdot \mathbf{D}_j(\mathbf{r})
 \right) + R_{ij}^{(0)} \left(  \mathbf{M}_i(\mathbf{r}) \cdot \mathbf{M}_j(\mathbf{r})
 \right)
+ } \\ \\
{\displaystyle
+ S_{ij}^{(0)} \left(  \mathbf{D}_i(\mathbf{r}) \cdot \mathbf{M}_j(\mathbf{r})
 \right) \biggr] \, n_i(\mathbf{r})\,  n_j(\mathbf{r}) +  }\\ \\
{\displaystyle + \sum_{i=1}^m \biggl[ \left( \mathbf{E}(\mathbf{r}) \cdot
\mathbf{D}_i(\mathbf{r}) \right) n_i(\mathbf{r}) + \left( \mathbf{H}(\mathbf{r}) \cdot
\mathbf{M}_i(\mathbf{r}) \right) n_i(\mathbf{r}) + \frac{1}{2} \left[ {\lambda_i(\mathbf{r})} \mathbf{D}_i^2 (\mathbf{r}) + {\nu_i(\mathbf{r})} \mathbf{M}_i^2 (\mathbf{r})  \right] \biggr] +}\\ \\
{\displaystyle + \sum_{i=0}^m \biggl[ T\, \ln
\left(\frac{n_i(\mathbf{r})}{n(\mathbf{r})}\right) -  \mu_i - \Psi(\mathbf{r})\, \omega_i \biggr]  n_i (\mathbf{r}).}
\end{array}
\end{equation}

The solution of the variational problem for functional~(\ref{lagr2}) is equivalent to solution of the variation problem for functional~(\ref{L2}) with conditions~(\ref{packing}), (\ref{N}), (\ref{D-M}). Thus, the equilibrium distributions of the components and their electric and magnetic moments obey the Lagrange-Euler system of equations for functional~(\ref{L3})
\begin{equation}\label{Lag-Eul}
    \frac{\partial \Lambda}{\partial u_s (\mathbf{r})}\ - \ \left(  \nabla\, \cdot  \frac{\partial \Lambda}{\partial \left(  \nabla u_s (\mathbf{r})  \right)} \right) \, = \, 0,
\end{equation}
together with conditions~(\ref{packing}), (\ref{N}), (\ref{D-M}).

Using the Lagrange-Euler equation for $u_s(\mathbf{r}) = n_i(\mathbf{r})  $  gives the equation:
\begin{equation}\label{L-ni}
\begin{array}{r}
     {\displaystyle  \sum_{j=1}^m \biggl[ K_{ij}^{(0)}
 + Q_{ij}^{(0)} \left(  \mathbf{D}_i(\mathbf{r})\cdot \mathbf{D}_j(\mathbf{r})
 \right) + R_{ij}^{(0)} \left(  \mathbf{M}_i(\mathbf{r}) \cdot \mathbf{M}_j(\mathbf{r})
 \right)
+ S_{ij}^{(0)} \left(  \mathbf{D}_i(\mathbf{r}) \cdot \mathbf{M}_j(\mathbf{r})
 \right) \biggr] \,  n_j(\mathbf{r})\, +} \\
{\displaystyle + \sum_{j=1}^m \sum_{\alpha} \biggl[  Q_{ij}^{(2)} \left( \nabla   {D}_i^{\alpha}(\mathbf{r})\cdot \nabla \left[ {D}_j^{\alpha}(\mathbf{r}) n_j(\mathbf{r}) \right] \right) + R_{ij}^{(2)} \left( \nabla   {M}_i^{\alpha}(\mathbf{r})\cdot \nabla \left[ {M}_j^{\alpha}(\mathbf{r}) n_j(\mathbf{r}) \right] \right)+ } \\
{\displaystyle
+ \frac{1}{2} \left[  S_{ij}^{(2)} \left( \nabla   {D}_i^{\alpha}(\mathbf{r})\cdot \nabla \left[ {M}_j^{\alpha}(\mathbf{r}) n_j(\mathbf{r}) \right] \right) +  S_{ij}^{(2)} \left( \nabla   {M}_i^{\alpha}(\mathbf{r})\cdot \nabla \left[ {D}_j^{\alpha}(\mathbf{r}) n_j(\mathbf{r}) \right] \right)  \right] \biggr] \, +}\\
{\displaystyle + \biggl[ \left( \mathbf{E}(\mathbf{r}) \cdot
\mathbf{D}_i(\mathbf{r}) \right)  + \left( \mathbf{H}(\mathbf{r}) \cdot
\mathbf{M}_i(\mathbf{r}) \right) + T\, \ln
\left(\frac{n_i(\mathbf{r})}{n(\mathbf{r})}\right) \biggr] -\mu_i + \omega_i\Psi \left( \mathbf{r} \right) = }\\
{\displaystyle = - \frac{1}{6} \sum_{j=1}^m \Biggl[   K_{ij}^{(2)} \Delta n_j (\mathbf{r}) + \sum_{\alpha} Q_{ij}^{(2)} \left(\nabla \cdot \left\{ D_i^{\alpha}(\mathbf{r})\,\nabla \left(D_j^{\alpha}(\mathbf{r})\, n_j (\mathbf{r})  \right)  \right\}  \right) +  }\\
{\displaystyle + \sum_{\alpha} R_{ij}^{(2)} \left(\nabla \cdot \left\{ M_i^{\alpha}(\mathbf{r})\,\nabla \left(M_j^{\alpha}(\mathbf{r})\, n_j (\mathbf{r})  \right)  \right\}  \right) +  }\\
{\displaystyle + \frac{1}{2} \sum_{\alpha} S_{ij}^{(2)} \biggl\{ \left(\nabla \cdot \left\{ D_i^{\alpha}(\mathbf{r})\,\nabla \left(M_j^{\alpha}(\mathbf{r})\, n_j (\mathbf{r})  \right)  \right\}  \right) + \left(\nabla \cdot \left\{ M_i^{\alpha}(\mathbf{r})\,\nabla \left(D_j^{\alpha}(\mathbf{r})\, n_j (\mathbf{r})  \right)  \right\}  \right)  \biggr\} \Biggr]};
\end{array}
\end{equation}

Using the Lagrange-Euler equation for $u_s(\mathbf{r}) = D_i^{\alpha}(\mathbf{r})$  gives the equation:
\begin{equation}\label{L-D}
\begin{array}{r}
{\displaystyle \sum_{j=1}^m \left\{ Q_{ij}^{(0)}\, D_j^{\alpha}(\mathbf{r}) + \frac{1}{2}\, S_{ij}^{(0)}\, M_j^{\alpha}(\mathbf{r})  \right\} n_i(\mathbf{r})\, n_j(\mathbf{r}) + E^{\alpha}(\mathbf{r})\, n_i(\mathbf{r}) + \lambda_i(\mathbf{r}) D_i^{\alpha}(\mathbf{r}) - }\\
{\displaystyle -\frac{1}{6}\, \sum_{j=1}^m \left\{Q_{ij}^{(2)} \left( \nabla n_i(\mathbf{r}) \cdot \nabla \left[D_j^{\alpha}(\mathbf{r}) n_j(\mathbf{r})   \right] \right) + \frac{1}{2}\, S_{ij}^{(2)} \left( \nabla n_i(\mathbf{r}) \cdot \nabla \left[M_j^{\alpha}(\mathbf{r}) n_j(\mathbf{r})   \right] \right)  \right\} =} \\
{\displaystyle = - \frac{1}{6}\, \nabla\cdot \sum_{j=1}^m \left\{ Q_{ij}^{(2)} n_i(\mathbf{r})\, \nabla \left[ D_j^{\alpha} (\mathbf{r}) n_j(\mathbf{r}) \right]  + \frac{1}{2}\,S_{ij}^{(2)} n_i(\mathbf{r})\, \nabla \left[ M_j^{\alpha} (\mathbf{r}) n_j(\mathbf{r}) \right]  \right\} };
\end{array}
\end{equation}

Similarly, using the Lagrange-Euler equation for $u_s(\mathbf{r}) = M_i^{\alpha}(\mathbf{r})$  gives the equation:
\begin{equation}\label{L-M}
\begin{array}{r}
{\displaystyle \sum_{j=1}^m \left\{ R_{ij}^{(0)}\, M_j^{\alpha}(\mathbf{r}) + \frac{1}{2}\, S_{ij}^{(0)}\, D_j^{\alpha}(\mathbf{r})  \right\} n_i(\mathbf{r})\, n_j(\mathbf{r}) + H^{\alpha}(\mathbf{r})\, n_i(\mathbf{r}) + \nu_i(\mathbf{r}) M_i^{\alpha}(\mathbf{r}) - }\\
{\displaystyle -\frac{1}{6}\, \sum_{j=1}^m \left\{R_{ij}^{(2)} \left( \nabla n_i(\mathbf{r}) \cdot \nabla \left[M_j^{\alpha}(\mathbf{r}) n_j(\mathbf{r})   \right] \right) + \frac{1}{2}\, S_{ij}^{(2)} \left( \nabla n_i(\mathbf{r}) \cdot \nabla \left[D_j^{\alpha}(\mathbf{r}) n_j(\mathbf{r})   \right] \right)  \right\} =} \\
{\displaystyle = - \frac{1}{6}\, \nabla\cdot \sum_{j=1}^m \left\{ R_{ij}^{(2)} n_i(\mathbf{r})\, \nabla \left[ M_j^{\alpha} (\mathbf{r}) n_j(\mathbf{r}) \right]  + \frac{1}{2}\,S_{ij}^{(2)} n_i(\mathbf{r})\, \nabla \left[ D_j^{\alpha} (\mathbf{r}) n_j(\mathbf{r}) \right]  \right\} }.
\end{array}
\end{equation}

The system of equations (\ref{L-ni}, \ref{L-D}, \ref{L-M}) with conditions~(\ref{packing}, \ref{N}, \ref{D-M}) describes space distributions of the components and local electric and magnetic moments for the case of full thermodynamic equilibrium in the system.

\subsection{Partial equilibrium in the system}

It should be noted that in view of colossal times of relaxations for solid state systems, the full thermodynamic equilibrium as a rule does not realizable. In the best way, there are the particular equilibriums in real solid state structures.

In particular, for inhomogeneous layered systems with electric and magnetic degrees of freedom relaxation times related to electric and magnetic degrees of freedom are short whereas times of redistribution of the components have a  colossal scales. Thus, the order parameters related to slow degrees of freedom, are not equilibrium parameters. They predetermine by the prehistory of the samples and should be preassigned.

Suppose that the slow variables in the system are space distributions of the components $n_i(\mathbf{r})$, whereas distributions of the electric $\mathbf{D}_i(\mathbf{r})$ and magnetic $\mathbf{M}_i(\mathbf{r})$ moments are the fast variables. Then the functions $n_i(\mathbf{r})$ should be prescribed, but the functions $\mathbf{D}_i(\mathbf{r})$ and $\mathbf{M}_i(\mathbf{r})$ obey the equations (\ref{L-D}, \ref{L-M}) and ~(\ref{packing}, \ref{N}, \ref{D-M}).

\section{Conclusion}

This paper sets as a goal development the unified theoretical approach to both statistical thermodynamics and kinetic phenomena in microheterogeneous systems with account their most essential real peculiarities:
\begin{enumerate}
    \item Presence of some inner degrees of freedom of constituent atoms (such as their atomic electric and magnetic moments);
    \item Presence of the competitive short-range and long-range parts of the components interatomic potentials;
    \item Presence of essential difference of the components atomic sizes. In particular, owing to this difference there are no ideal lattices in the real condensed matter.
\end{enumerate}

It should be noted that such peculiarities as misfits on the phases boundaries, deformations fields, and local elastic properties of the matter in microheterogeneous systems are caused mainly by the interatomic potentials. Therefore, the correct account of the indicated peculiarities of the real systems should be as basis for the adequate model of the real microheterogeneous systems, especially for the layered and composite systems with essential magnetoelectric interaction.

There are many works dealing with theoretical research of microheterogeneous systems with the magnetoelectric effect. The reviews of some results on the subject my be find in papers~\cite{Bich,Rabe,Bich2,Bich3}. Most of the theoretical papers are based on the phenomenological Ginzburg-Landau-like free energy functional with additional terms considering  lattices distortions and misfits on the interphase boundaries~\cite{Per,Chen}. There is a series of the papers by Khachaturyan with co-workers~\cite{Khach1,Khach2} based on the phase field approach to composites. As a rule the nature of the phenomenological parameters remains hidden.

It should be noted the very interesting works on the statistical properties and random geometry of inhomogeneous systems~\cite{georgii}.

The present paper contains the other approach to statistical thermodynamics of the multicomponent condensed matter. The main peculiarities of this approach consist of following.
\begin{enumerate}
    \item Instead of the ideal lattices we introduce the packing condition~(\ref{packing}). This condition allows to take into account essential differences of the components atomic sizes.
    \item We take into account the long-range parts of the interatomic potentials by means of the effective fields approximation. Owing to packing condition the GLM as a whole falls outside the limits of the effective fields approximation.
    \item We take into account the interatomic potentials depending on inner degrees of freedoms such as atomic electric and/or magnetic momentum. The magnetoelectric interactions are included in the implicit form.
    \item The Helmholtz free energy functional for the generalized lattice model with account of electric and magnetic atomic moments is reduced to GLCH-like form.  The connection between the interatomic potentials and parameters of the GLCH-like functional is established.
\end{enumerate}

One of the most interesting applications of the GLM is research of layered structures with alternation of magnetic and ferroelectric layers. In particular, this approach permits
\begin{itemize}
    \item to take into account lattices misfits on the interphases without any auxiliary assumptions;
    \item to impart the physical interpretation to the phenomenological models like GLCH;
    \item to find the ways for prognosis of the layered structures.
\end{itemize}
These results will be presented in the next papers.

\section*{Acknowledgments}

The work was partially supported by the Program of Russian Ministry of Education and Science.

\end{document}